\documentclass{appolb}
\usepackage{epsfig}
\usepackage{amssymb}
\usepackage{color}

\begin{document}

\pagestyle{plain}

\title{Different fractal properties of positive and negative returns}

\author{P.~O\'swi\c ecimka$^1$, J.~Kwapie\'n$^1$, S.~Dro\.zd\.z$^{1,2}$, 
A. Z.~G\.orski$^1$, R. Rak$^2$ \address{$^1$Institute of Nuclear Physics, 
Polish Academy of Sciences, \\ PL--31-342 Krak\'ow, Poland\\ $^2$Institute 
of Physics, University of Rzesz\'ow, PL--35-310 Rzesz\'ow, Poland}}

\maketitle

\begin{abstract} 
We perform an analysis of fractal properties of the positive and the 
negative changes of the German DAX30 index separately using Multifractal 
Detrended Fluctuation Analysis (MFDFA). By calculating the singularity 
spectra $f(\alpha)$ we show that returns of both signs reveal 
multiscaling. Curiously, these spectra display a significant difference in 
the scaling properties of returns with opposite sign. The negative price 
changes are ruled by stronger temporal correlations than the positive 
ones, what is manifested by larger values of the corresponding H\"{o}lder 
exponents. As regards the properties of dominant trends, a bear market is 
more persistent than the bull market irrespective of the sign of 
fluctuations. 
\end{abstract}
 

\section{Introduction}

Typical signals generated by economic systems are non-trivial structures 
which can be characterized in terms of the theory of multifractals. 
Interestingly, these structures are to some degree universal in real 
world, since they come not only from finance but also from diverse fields 
of science like physics~\cite{Monthus2007,schaefer2006,%
nogueira2006,ordemann2006,barthelemy2000}, chemistry or 
biology~\cite{peng1994,buldyrev1995,arneodo1996,hausdorff2001}. The 
concept of "fractal world" was proposed by Mandelbrot in 1980s and was 
based on scale-invariant statistics with power law 
correlations~\cite{mandelbrot1982}. In subsequent years this new theory 
was developed and finaly it brought a more general concept of 
multiscaling. It allows one to study the global and local behaviour of 
a singular measure, or, in other words, the mono- and multifractal 
properties of a system. In economy, mutifractality is a one of the well 
known stylized facts which characterize non-trivial properies of financial 
time series~\cite{eisler2004}. The stock price (or index) fluctuations can 
be described in terms of long-range temporal correlations by a spectrum of 
the H\"{o}lder-Hurst exponents and a set of fractal dimensions. Obtained 
results show that there exist n-point correlations in financial data, 
hardly detectable with commonly used methods like power spectrum or 
autocorrelation function. This discovery allows us to reject the efficient 
market hipothesis (EMH) with its main assumption that retuns are 
uncorrelated. Of course this kind of analysis is possible because 
appropriate methods were developed in last decade, among which the most 
popular are Wavelet Transform Modulus Maxima (WTMM) and 
Multifractal Detrended Fluctuation Analysis (MFDFA). As one of our recent 
works proved~\cite{oswiecimka2006}, the latter method is more reliable when 
the fractal properties of the analyzed signals are not known {\it a 
priori} and this is why we prefer to use this method here.

In a standard approach, one assumes that both the positive and the 
negative fluctuations have the same fractal or scaling properties; 
however, this may not apply to some particular cases~\cite{ohashi2003}. 
For example, studying deeper characteristics of the financial signals we 
can infer that the nature of fluctuations can depend on their 
direction~\cite{ferreira2007}. Therefore, in order to apprehend the 
studied processes completely we have to take into consideration also their 
sign. This is a reason why we decided to generalize MFDFA, to be able to 
analyze the positive and the negative changes separately.

This paper is organized as follows. In Section 2, we describe the data and 
explain the method in detail. Section 3 presents the results and 
disscusion and, finally, section 4 concludes.

\section{Data and Methodology}

All the calculations were performed for high-frequency data from the 
German stock market index DAX, comprising the two following periods: 
Period 1 from Nov 28, 1997 to Dec 30, 1999 and Period 2 from May 1, 2002 
to May 1 2004. The time interval between consecutive records was $\Delta t 
= 1$min. In each case the logarithmic returns were calculated: 
$g(i)=\ln(p(t_{i}+\Delta t))-\ln(p(t_{i}))$, where $p(t_{i})$ denotes an 
index value in a moment $t_{i}$. In addition, we removed all the overnight 
returns, because they cover a much longer time interval than 1 min and are 
also contaminated by some spurious artificial effects~\cite{gorski}. The 
length of the time series was approximately 268,000 points and it was 
enough to obtain statisticaly significant results. Moreover, we also 
analized two shorter time series (from Nov 28, 1997 to July 15, 1998 and from July 16 to Oct 15, 1998) which represent the periods of a bull and a bear market, respectively.

In order to investigate the fractal properties of the positive and the 
negative index fluctuations separetely, we modified the algorythm of
MFDFA~\cite{kantelhart} such that the natural scale of signal and the length of possible temporal correlations is preserved. The main steps of this procedure can be briefly 
sketched as follows. At first one divides a given time series $g(i)$ 
into $M_{s}$ disjoint segments of length $s$ starting from the begining 
of the $g(i)$. To avoid neglecting the data which don't fall into any 
segment (it refers to the data at the end of $g(i)$) the procedure is 
repeated starting this time from the end of the time series. Finally, one 
has $2M_{s}$ segments total. For each segment $\nu$, two signal profiles 
have to be calculated, separately for the positive (\textit{p}) and the negative (\textit{n}) 
fluctuations:
\begin{equation}
Y_{p}^{\nu}(i,s)=\Sigma_{k=1}^{i}g(Q^{\nu}(k)) \quad i=1,...,N_{p}^{\nu} 
\end{equation}
\begin{equation}
Y_{n}^{\nu}(j,s)=\Sigma_{l=1}^{j}g(R^{\nu}(l)) \quad j=1,...,N_{n}^{\nu},
\end{equation}
where $Q^{\nu}(k)$ and $R^{\nu}(l)$ denote the sets of ($N^{\nu}_{p(n)}$) positions of the 
positive and the negative returns, respectively, within a segment $\nu$. 
In the next step we evaluate the variance for each segment:
\begin{equation}
F^{2}_{p}(\nu,s)=(\frac{1}{s}\Sigma_{k=1}^{N_{p}^{\nu}}\lbrace Y_{p}^{\nu}(k,s)-P^{l}_{\nu}(k)\rbrace)
\end{equation}
and
\begin{equation}
 F^{2}_{n}(\nu,s)=(\frac{1}{s}\Sigma_{l=1}^{N_{n}^{\nu}}\lbrace 
Y_{n}^{\nu}(l,s)-P^{l}_{\nu}(l)\rbrace),
\end{equation}
where $P^{l}_{\nu}()$ is a local trend in a segment $\nu$; it can be 
approximated by fitting an $l$th order polynomial $P^{l}_{\nu}$. This 
trend has to be substracted from the data. In this paper we use $l=2$ so we can eliminate \textit{l} order possible trend in the profile and \textit{l-1} in the original time series.
By averaging $F^{2}_{p}(\nu,s)$ and $ F^{2}_{n}(\nu,s)$ over all $\nu$'s 
we obtain the $q$th-order fluctuation functions:
\begin{equation}
F_{p}^{q}(s)=\lbrace\frac{1}{2M_{s}}\Sigma_{\nu-1}^{2M_{s}}[F^{2}_{p}(\nu,s)]^{q/2}\rbrace^{1/q}
\end{equation}
\begin{equation}
F_{n}^{q}(s)=\lbrace\frac{1}{2M_{s}}\Sigma_{\nu-1}^{2M_{s}}[F^{2}_{n}(\nu,s)]^{q/2}\rbrace^{1/q},
\end{equation}
where $q\in\Re$ (in this paper, to make the results more readable, we use 
$-10<q<10$~\cite{oswiecimka2005}). Of course, this procedure has to be 
repeated for different segment lengths $s$. For a signal with fractal 
properies the fluctuation functions reveal power-law scaling
\begin{equation}
F_{p(n)}^{q}(s) \sim s^{h_{p(n)}(q)} 
\end{equation}
for large $s$. Family of the generalized Hurst exponents $h(q)$ 
characterizes complexity of an analyzed fractal. For a monofractal signal 
$h(q)=const$, while for multifractal signals $h(q)$ is a decreasing 
function of $q$. By knowing the spectrum of the generalized Hurst 
exponents for fluctuations with different signs we are able to calculate 
the singularity spectrum $f_{p(n)}(\alpha)$ according to the following 
relations:
\begin{equation}
\tau(q)=qh(q)-1
\end{equation}
\begin{equation}
\alpha=\tau'(q)\quad \textrm{and} \quad f(\alpha)=q\alpha-\tau(q),
\end{equation}
where $\alpha$ is called the singularity exponent and $f(\alpha)$ is a 
fractal dimension of the set of all points $x_{0}$ such that 
$\alpha(x_0)=\alpha$.
 
\section{Results}

\begin{figure}
\begin{center}
\includegraphics[width=0.7\textwidth,height=0.55\textwidth]{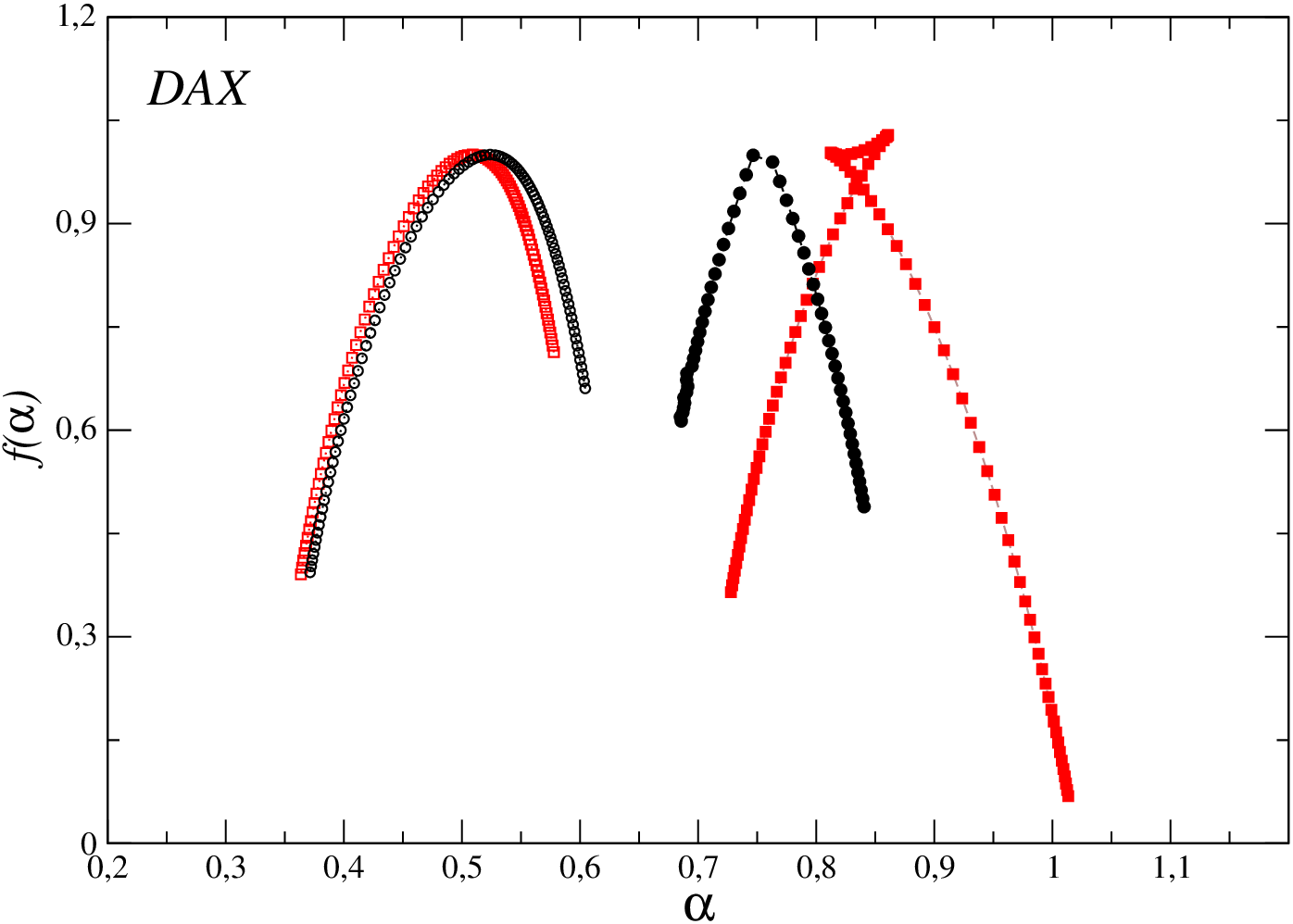}
\caption{Singularity spectra for negative (red squares) and positive (black circles) DAX
returns from Period 1 (Dec 1997 $-$ Dec 1999). Closed symbols refer to 
original and open to shuffled times series.}
\label{DAX_98_99}
\end{center}
\end{figure}
\begin{figure}
\begin{center}
\includegraphics[width=0.7\textwidth,height=0.55\textwidth]{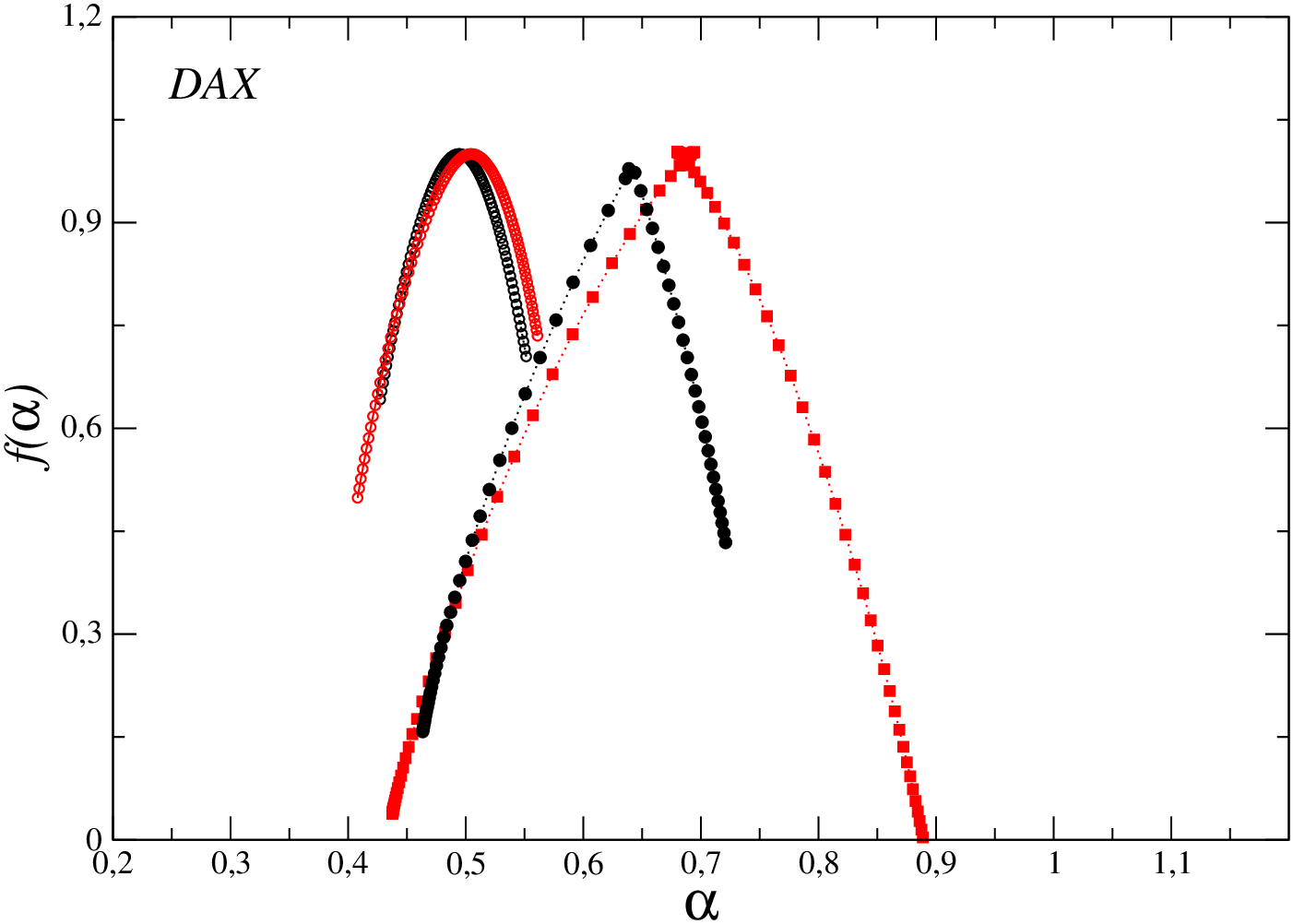}
\caption{Singularity spectra for negative (red squares) and positive (black circles) DAX
returns in Period 2 (May 2004 $-$ May 2006). Filled symbols refer to 
original and open to shuffled times series.}
\label{DAX_04_04}
\end{center}
\end{figure}

Figure \ref{DAX_98_99} presents the $f_{p}(\alpha)$ and $f_{n}(\alpha)$ 
spectra for DAX in Period 1. It is easily visible that these spectra 
are different. For the negative fluctuations $f_{n}(\alpha)$ is rather 
wide ($\Delta \alpha \approx 0.3$) with its maximum placed at $\alpha^{max}_{n} 
\approx 0.85$. $f_{p}(\alpha)$ is much narrower ($\Delta \alpha \approx 
0.15$) than in the former case; its maximum corresponds to $\alpha^{max}_{p} 
\approx 0.73$. In both cases, the positions of the maxima indicate a 
persistent character of the related index fluctuations. Naturally, if one
looks at the scaling properties of volatility, one can expect such 
behaviour, but the shift between $f_{p}(\alpha)$ and $f_{n}(\alpha)$ 
as well as the difference in the spectra widths is a completely new 
observation. The $f_{n}(\alpha)$ is wider than its conterpart for the 
positive returns, suggesting that a richer multifractal (or more 
complex dynamics) is seen for the negative fluctuactions. For the shuffled 
signals, properties of the singularity spectrum do not depend on 
a direction of index changes. A lack of temporal correlations is 
manifested by a position of the spectrum at $\alpha^{max}_{p,n} \approx 0.5$. The 
difference between $f_{p}(\alpha)$ and $f_{n}(\alpha)$ in this case is rather 
meaningless and is a consequence of a finite sample size. Similar results 
we can see in Figure \ref{DAX_04_04} (Period 2). Again, the 
$f_{n}(\alpha)$ is shifted to the right (maximum at $\alpha^{max}_{n} \approx 
0.7$) relative to the spectrum for the positive returns ($\alpha^{max}_{p} 
\approx 0.65$); however, the difference is rather small in this case. 
Moreover, the multifractal spectrum for the negative index changes is 
substantially wider ($\Delta \alpha \approx 0.45$) than for 
$f_{p}(\alpha)$ ($\Delta \alpha \approx 0.25$) and this indicates a more 
complex dynamics governing behaviour of the negative returns. For the 
mixed-up data the spectra look almost identically with their maximum at 
$\alpha^{max}_{n,p} \approx 0.5$.

The multifractal characteristics of data can depend on a considered 
timeframe~\cite{oswiecimka2006APPB}. In particular, the multifractal 
spectrum can evolve in time to reflect the changing scaling properties of 
the data under study. In order to investigate how different market 
phases, associated with different behaviour of investors, can manifest 
themselves in the singularity spectra of the index returns, we applied our 
method to the bull and the bear phases, separately.
\begin{figure}
\begin{center}
\includegraphics[width=0.75\textwidth,height=0.65\textwidth]{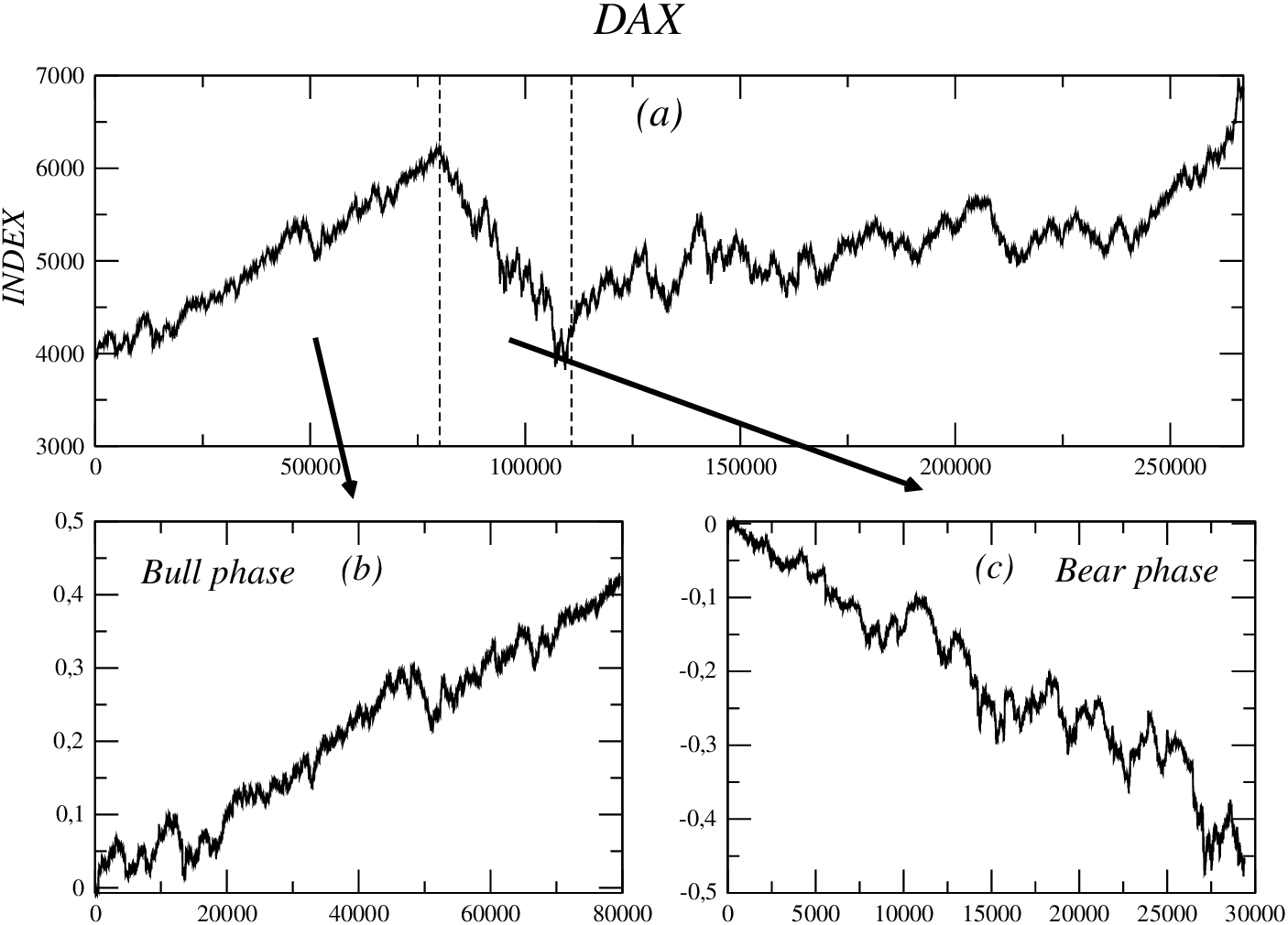}
\caption{DAX daily closings in Period 1 (a) and the zoomed subperiods of 
index rise (b) and index decline (c).}
\label{boom_and slump}
\end{center}
\end{figure}
Figure \ref{boom_and slump} shows the intervals of persistent growth and 
sudden decrease of the DAX index during Period 1. Results of our fractal 
analysis for these two intervals are presented in Figure \ref{boom_and 
slump_results}.
There is a clear difference between spectra for the growth and 
the decrease phase. For the period of slump the singularity spectra are 
shifted to the right, what means stronger correlations (both for 
the negative and the positive changes) than in case of boom. The position 
of maximum for the negative fluctuations is localized approximatetly at 
$\alpha^{max}_{n} \approx 0.87$ for the bear phase, whereas for the bull phase 
the maximum is placed at $\alpha^{max}_{p} \approx 0.82$; this gives the 
discrepancy $\Delta \alpha \approx 0.2$. For the positive fluctuations the 
difference is even more apparent and it totals $\Delta\alpha \approx 
0.25$. By analyzing these relations between the spectra for the returns of 
different sign we can formulate a conclusion that the negative 
fluctuations are more persistent (or stronger correlated) than series of the opposite sign. This 
phanomenon is reflected in positions of the maxima of $f(\alpha)$ (higher 
$\alpha^{max}_{n}$). 
\begin{figure}
\begin{center}
\includegraphics[width=0.7\textwidth,height=0.55\textwidth]{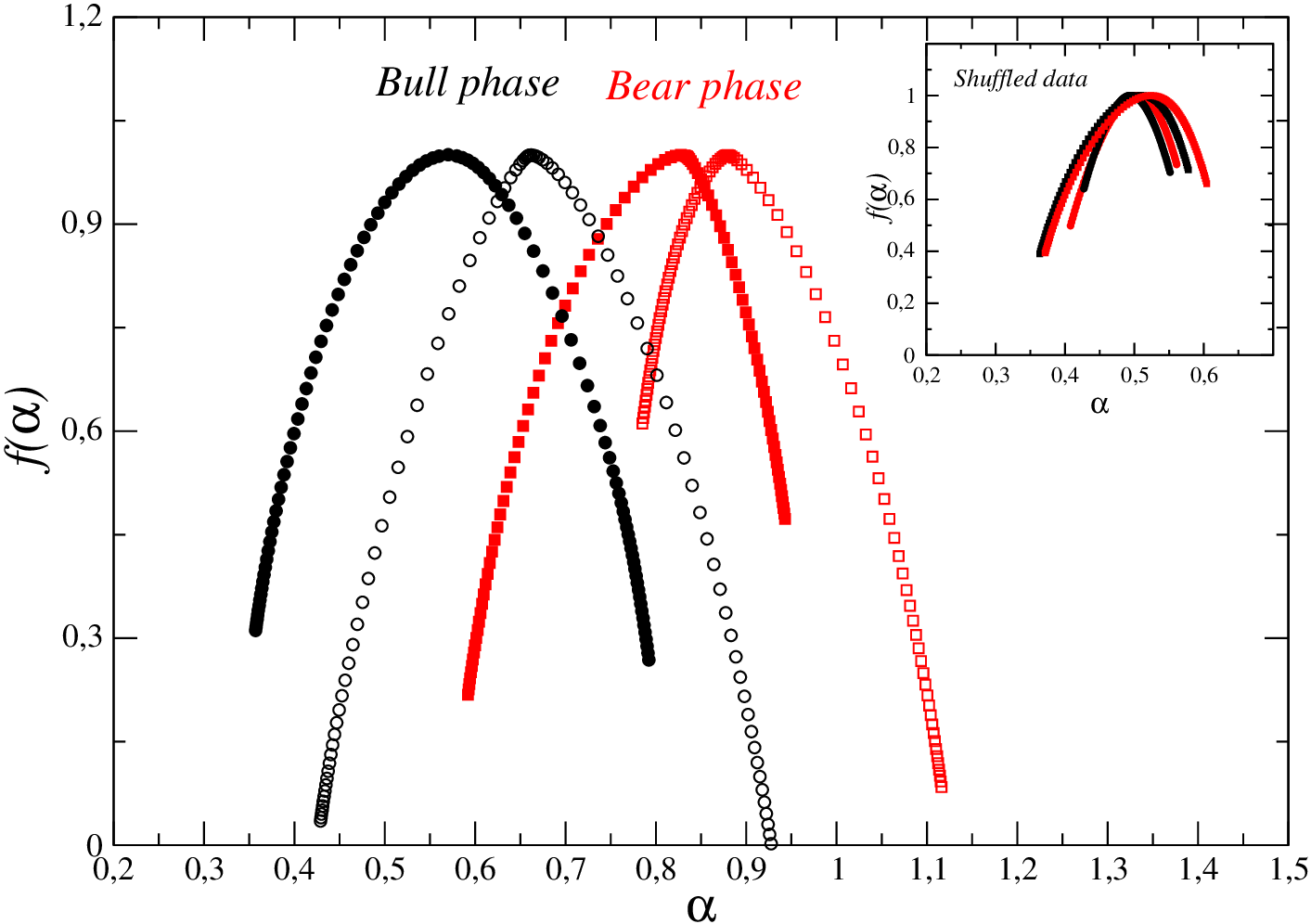}
\caption{Comparison of $f(\alpha)$ spectra for phase of growth (black circles) 
and phase of decrease (red squares). Open symbols refer to negative and filled 
to positive DAX fluctuations.}
\label{boom_and slump_results}
\end{center}
\end{figure}
The width of the singularity 
spectra for the bear phase is $\Delta \alpha \approx 0.35$ irrespective of 
a sign. For the bull phase, on the other hand, the $f(\alpha)$ spectrum is 
wider for the positive changes ($\Delta \alpha \approx 0.45$) than 
for the negative ones $\Delta \alpha \approx 0.35$; it shows richer 
multifractality in the former case. For the shuffed series the spectra 
have approximately the same width $\Delta \alpha \approx 0.2$ and are 
localized in a close vicinity of $\alpha \approx 0.5$. This demostrates 
that the temporal correlations present in time series are responsible for
the discrepancy in fractal properties between the bull and bear phases.

\section{Conclutions}

We applied the MFDFA technique to show a difference in the fractal 
properties of the negative and the positive DAX index fluctuations. Our 
results suggest that a more persistent behaviour and often richer 
multifractality is associated with the negative price changes. This 
asymmetry disappears for the shuffled signals what implies that 
the temporal correlations are solely responsible for this effect. 
Moreover, our study of the index trends indicates a significant 
discrepancy between the bear and the bull market. Declining market 
is much more correlated than the rising one and can be described in terms 
of the H\"{o}lder exponent by $\alpha$ close to 1. We believe that 
the asymmetric fractal properties can give us an opportunity to better 
understand the mechanism that governs the stock market dynamics. From a 
practical point of view this fact can have applications in modeling and 
forecasting the stock market data and may be an important factor in risk 
evaluation.


\begin{thebibliography}{99}
\bibitem{Monthus2007} C. Monthus, T. Garel, Phys. Rev. E \textbf{75}, 051122 (2007).
\bibitem{schaefer2006} B.M. Sch\"{a}fer, W. Hofmann, H. Lampeitl, M. Hemberger, Nucl.Instrum.Meth. A \textbf{465}, 394 (2001).
\bibitem{nogueira2006} E. Nogueira Jr., R. F. S. Andrade, S. Coutinho, Physica A textbf{360}, 365 (2006).
\bibitem{ordemann2006} A. Ordemann, M. Porto, H. E. Roman, S. Havlin, A. Bunde, Phys. Rev. E \textbf{61}, 6858 (2000),
\bibitem{barthelemy2000} M. Barthelemy, S.V. Buldyrev, S. Havlin, H.E. Stanley, Phys. Rev. E (Rapid Comm.) \textit{61}, R3283 (2000).
\bibitem{peng1994} C.-K. Peng, S. V. Buldyrev, S. Havlin, M. Simons, H. E. Stanley, and A. L. Goldberger, Phys. Rev E \textbf{49}, 1685 (1994).
\bibitem{buldyrev1995} S. V. Buldyrev, A. L. Goldberger, S. Havlin, R. N. Mantegna, M. E. Matsa, C.-K. Peng, M. Simons, and H. E. Stanley, Phys. Rev. E \textbf{51}, 5084 (1995).
\bibitem{arneodo1996} A.Arneodo, Y. d'Aubenton-Carafa, E. Bacry, P.V. Graves, J.F.  Muzy, C. Thermes, Physica D \textit{96}, 291 (1996).
\bibitem{hausdorff2001} J.M. Hausdorff,Y. Ashkenazy,C.-K. Peng, P.Ch. Ivanov, H.E. Stanley and A.L. Goldberger, Physica A \textbf{302}, 138 (2001).
\bibitem{mandelbrot1982} B.B. Mandelbrot, W.H. FreeMan, New York (1982).
\bibitem{eisler2004} Z. Eisler, J.Kert\'{e}sz, Physica A \textbf{343}, 603 (2004).
\bibitem{ohashi2003} K. Ohashi, L.A.N. Amaral, B.H. Natelson and Y. Yamamoto, Phys. Rev. E \textbf{68}, 065204(R) (2003).
\bibitem{ferreira2007} N.B. Ferreira, R. Menezes, D.A. Mendes,  Physica A \textbf{382}, 73 (2007).
\bibitem{gorski} A.Z. G\'{o}rski, S. Dro\.{z}d\.{z}, J. Speth, Physica A \textbf{316}, 496 (2002) .
\bibitem{kantelhart} J.W. Kantelhardt, S.A. Zschiegner, E. Koscielny-Bunde, S. Havlin, A. Bunde, H.E. Stanley, Physica A \textbf{316}, 87 (2002).
\bibitem{oswiecimka2006} P. O\'{s}wi\c ecimka, J. Kwapie\'{n} and Stanisaw Dro\.{z}d\.{z}, Phys. Rev. E \textbf{74}, 016103 (2006).
\bibitem{oswiecimka2006APPB} P. O\'{s}wi\c ecimka, J. Kwapie\'{n}, S. Dro\.{z}d\.{z}, A.Z. G\'{o}rski, R. Rak, Acta Phys. Polon. B \textbf{37}, 3083 (2006).
\bibitem{oswiecimka2005} P. O\'{s}wi\c ecimka, J. Kwapie\'{n}, S. Dro\.{z}d\.{z}, Physica A \textbf{347}, 626 (2005).
\end{thebibliography}
\end{document}